\documentclass[aps,preprintnumbers,tightenlines,12pt,nofootinbib,superscriptaddress,amsmath,amssymb]{revtex4}

\usepackage{graphicx}
\usepackage{bm}
\usepackage{textcomp}

\begin{document}

\preprint{Published in JINST {\bf 4} P10010 (2009)}

\title{A novel camera type for very high energy gamma-ray astronomy based on
  Geiger-mode avalanche photodiodes}\thanks{This is an author-created,
  un-copyedited version of an article published in Journal of
  Instrumentation. IOP Publishing Ltd is not responsible for any errors or
  omissions in this version of the manuscript or any version derived from
  it. The definitive publisher authenticated version is available online at
  http://www.iop.org/EJ/abstract/1748-0221/4/10/P10010.\\}

\author{H.~Anderhub}
\affiliation{Institute for Particle Physics, ETH Zurich,\\
  Schafmattstr.~20, 8093 Zurich, Switzerland}
\author{M.~Backes}
\affiliation{TU Dortmund University,\\
  Otto-Hahn-Str.~4, 44227 Dortmund, Germany}
\author{A.~Biland}
\affiliation{Institute for Particle Physics, ETH Zurich,\\
  Schafmattstr.~20, 8093 Zurich, Switzerland}
\author{A.~Boller}
\affiliation{Institute for Particle Physics, ETH Zurich,\\
  Schafmattstr.~20, 8093 Zurich, Switzerland}
\author{I.~Braun}
\affiliation{Institute for Particle Physics, ETH Zurich,\\
  Schafmattstr.~20, 8093 Zurich, Switzerland}
\author{T.~Bretz}
\affiliation{University of W\"urzburg,\\
  Am Hubland, 97074 W\"urzburg, Germany}
\author{S.~Commichau}
\affiliation{Institute for Particle Physics, ETH Zurich,\\
  Schafmattstr.~20, 8093 Zurich, Switzerland}
\author{V.~Commichau}
\affiliation{Institute for Particle Physics, ETH Zurich,\\
  Schafmattstr.~20, 8093 Zurich, Switzerland}
\author{D.~Dorner}
\affiliation{Institute for Particle Physics, ETH Zurich,\\
  Schafmattstr.~20, 8093 Zurich, Switzerland}
\affiliation{ISDC Data Center for Astrophysics, University of Geneva,\\
  Chemin d'Ecogia 16, 1290 Versoix, Switzerland}
\author{A.~Gendotti}
\affiliation{Institute for Particle Physics, ETH Zurich,\\
  Schafmattstr.~20, 8093 Zurich, Switzerland}
\author{O.~Grimm}
\affiliation{Institute for Particle Physics, ETH Zurich,\\
  Schafmattstr.~20, 8093 Zurich, Switzerland}
\author{H.~von Gunten}
\affiliation{Institute for Particle Physics, ETH Zurich,\\
  Schafmattstr.~20, 8093 Zurich, Switzerland}
\author{D.~Hildebrand}
\affiliation{Institute for Particle Physics, ETH Zurich,\\
  Schafmattstr.~20, 8093 Zurich, Switzerland}
\author{U.~Horisberger}
\affiliation{Institute for Particle Physics, ETH Zurich,\\
  Schafmattstr.~20, 8093 Zurich, Switzerland}
\author{T.~Kr\"ahenb\"uhl}
\affiliation{Institute for Particle Physics, ETH Zurich,\\
  Schafmattstr.~20, 8093 Zurich, Switzerland}
\author{D.~Kranich}
\affiliation{Institute for Particle Physics, ETH Zurich,\\
  Schafmattstr.~20, 8093 Zurich, Switzerland}
\author{E.~Lorenz}
\affiliation{Institute for Particle Physics, ETH Zurich,\\
  Schafmattstr.~20, 8093 Zurich, Switzerland}
\affiliation{Max Planck Institute for Physics,\\
  F\"ohringer Ring 6, 80805 Munich, Germany}
\author{W.~Lustermann}
\affiliation{Institute for Particle Physics, ETH Zurich,\\
  Schafmattstr.~20, 8093 Zurich, Switzerland}
\author{K.~Mannheim}
\affiliation{University of W\"urzburg,\\
  Am Hubland, 97074 W\"urzburg, Germany}
\author{D.~Neise}
\affiliation{TU Dortmund University,\\
  Otto-Hahn-Str.~4, 44227 Dortmund, Germany}
\author{F.~Pauss}
\affiliation{Institute for Particle Physics, ETH Zurich,\\
  Schafmattstr.~20, 8093 Zurich, Switzerland}
\author{D.~Renker}
\affiliation{Paul Scherrer Institute,\\
  5232 Villigen PSI, Switzerland}
\author{W.~Rhode}
\affiliation{TU Dortmund University,\\
  Otto-Hahn-Str.~4, 44227 Dortmund, Germany}
\author{M.~Rissi}
\affiliation{Institute for Particle Physics, ETH Zurich,\\
  Schafmattstr.~20, 8093 Zurich, Switzerland}
\author{U.~R\"oser}
\affiliation{Institute for Particle Physics, ETH Zurich,\\
  Schafmattstr.~20, 8093 Zurich, Switzerland}
\author{S.~Rollke}
\affiliation{TU Dortmund University,\\
  Otto-Hahn-Str.~4, 44227 Dortmund, Germany}
\author{L.~S.~Stark}
\affiliation{Institute for Particle Physics, ETH Zurich,\\
  Schafmattstr.~20, 8093 Zurich, Switzerland}
\author{J.-P.~Stucki}
\affiliation{Institute for Particle Physics, ETH Zurich,\\
  Schafmattstr.~20, 8093 Zurich, Switzerland}
\author{G.~Viertel}
\affiliation{Institute for Particle Physics, ETH Zurich,\\
  Schafmattstr.~20, 8093 Zurich, Switzerland}
\author{P.~Vogler}
\affiliation{Institute for Particle Physics, ETH Zurich,\\
  Schafmattstr.~20, 8093 Zurich, Switzerland}
\author{Q.~Weitzel}
\altaffiliation{Corresponding author}
\email{qweitzel@phys.ethz.ch}
\affiliation{Institute for Particle Physics, ETH Zurich,\\
  Schafmattstr.~20, 8093 Zurich, Switzerland}

\begin{abstract}
  \vspace{1cm}
  Geiger-mode avalanche photodiodes (G-APD) are promising new sensors for
  light detection in atmospheric Cherenkov telescopes. In this paper, the
  design and commissioning of a 36-pixel G-APD prototype camera is
  presented. The data acquisition is based on the Domino Ring Sampling (DRS2)
  chip. A sub-nanosecond time resolution has been achieved. Cosmic-ray induced
  air showers have been recorded using an imaging mirror setup, in a
  self-triggered mode. This is the first time that such measurements have been
  carried out with a complete G-APD camera.
\end{abstract}

\maketitle
\newpage

\section{Introduction}

Imaging atmospheric Cherenkov telescopes (IACT) have been very successful in
detecting very high energy ($\sim$ 0.1--30\,TeV) $\gamma$-rays from cosmic
sources \cite{aharonian08}. The key component is a pixelated camera which has
to resolve flashes of Cherenkov light from air showers (duration 1--5\,ns for
$\gamma$-induced air showers, main wavelength range
300--650\,nm). High-sensitivity photo-sensors are needed, even using
light-collecting optics, since e.\,g. a 1\,TeV primary photon hitting the
atmosphere only results in about one hundred Cherenkov photons per square
meter. Until now, matrices of photomultiplier tubes (PMT) have always been
employed for this task. This is a well-known technology which, however, comes
with some intrinsic disadvantages for telescope applications. PMTs are rather
heavy and bulky, but at the same time fragile. They require high voltages of
several 100\,V or even kV and are damaged when exposed to sunlight. Typical
PMTs furthermore have photon detection efficiencies (PDE) of only 20--30$\%$.

Since a few years, a new type of semiconductor light-sensor is being
developed, the so-called Geiger-mode avalanche photodiode (G-APD\footnote{Also
  referred to as silicon photomultiplier (SiPM) or multi-pixel photon counter
  (MPPC).}) \cite{renker09}. These light-weight devices are built up from
multiple APD cells operated in Geiger-mode. All the cells are connected in
parallel and the overall signal is the sum of all simultaneously fired
cells. They need bias voltages of 50--100\,V, usually 1--5\,V above the
breakdown voltage. A high gain similar to PMTs is reached ($10^5$--$10^6$)
and, potentially, higher PDEs of up to 50$\%$. The market for G-APDs is
continuously growing, and several manufacturers are working on their
improvement. For an IACT application under realistic ambient conditions,
several technical challenges have to be met. This mostly concerns the
necessity to compensate for gain variations due to changes in temperature or
night-sky background light (NSB). While the former is an intrinsic feature of
G-APDs, the latter is a general problem and applies to other photo-sensors as
well. An advantage of G-APDs is that they can be operated at NSB rates up to
several GHz per sensor, thus allowing measurements under twilight and
moonlight conditions.

First tests to detect Cherenkov light with small G-APD arrays have been
performed in the past \cite{biland08}. In order to develop a complete G-APD
camera and find solutions for the technical challenges, the FACT project
(First G-APD Camera Test) \cite{braun09} has been launched. The prototype
presented in this paper marks the first step towards a large camera with a
field of view of about $5^{\circ}$ (0.1--0.2$^{\circ}$ per pixel). This device
is foreseen for the DWARF telescope (Dedicated multi-Wavelength AGN Research
Facility) \cite{bretz08} which will perform monitoring of strong and varying
$\gamma$-ray sources.

\section{Camera layout and front-end electronics}

\begin{figure}[t]
  \begin{center}
    \includegraphics[width=.63\textwidth]{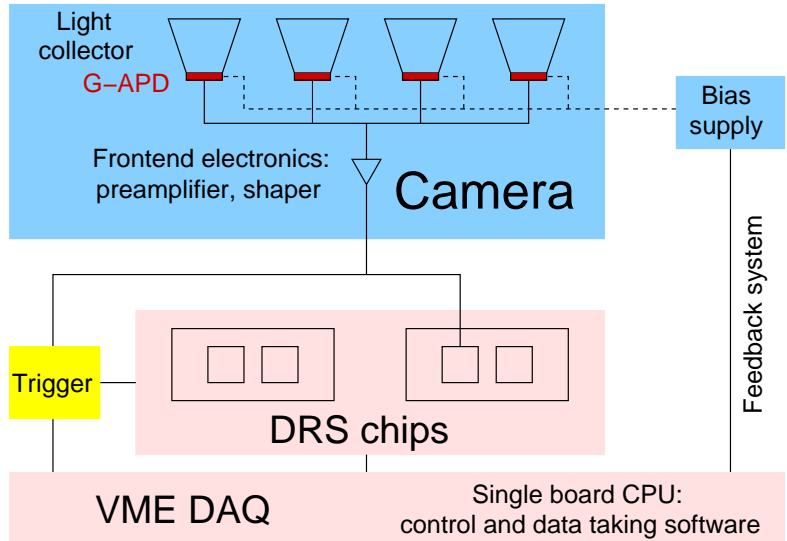}
    \caption{Schematic overview of the G-APD camera layout (upper part) and
      the DAQ system (lower part); signal flow of one pixel shown. The camera
      is mounted inside a water-tight box, while the DAQ and trigger
      components are located in a counting room.}
    \label{fig:schematic}
  \end{center}
\end{figure}

The prototype module comprises 144 G-APDs of type Hamamatsu MPPC
\mbox{S10362-33-50C}~\cite{hamamatsu08}. Each has a sensitive area of 3\,mm
$\times$ 3\,mm, covered by 3600 cells of 50\,\textmu m $\times$ 50\,\textmu m
size. The operating bias voltage ranges from 71.15\,V to 71.55\,V for the 144
sensors (at $25^{\circ}$C), corresponding to a gain of $7.5\cdot 10^5$. The
dark count rate is below 5\,MHz. A non-imaging light collector is placed on
top of each G-APD to compensate for dead spaces due to the diode-chip
packaging (see figure~\ref{fig:schematic}, top part). Open aluminum cones with
an affixed reflecting foil (ESR Vikuiti\texttrademark\,by 3M) and a quadratic
base are in use (bottom side: 2.8\,mm $\times$ 2.8\,mm, top side: 7.2\,mm
$\times$ 7.2\,mm, height: 17.5\,mm, effective solid angle: 0.55\,sr). With
such collectors, the NSB rate during the darkest nights at e.g. the
Observatorio Roque de los Muchachos, La Palma, is $\sim 10$\,MHz per sensor
\cite{mirzoyan94}.

\begin{figure}[t]
  \begin{center}
    \includegraphics[height=3.7cm]{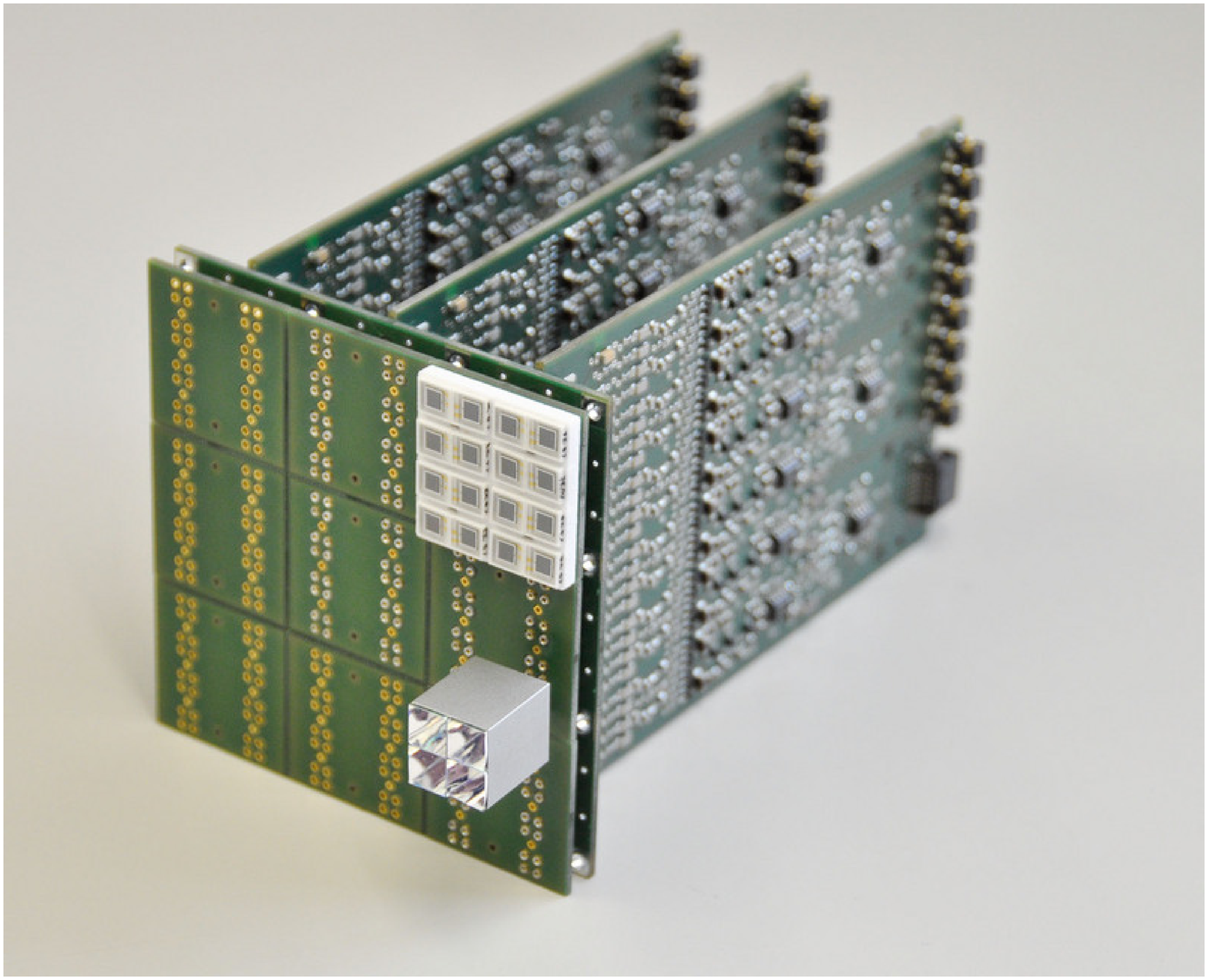}
    \hfill
    \includegraphics[height=3.7cm]{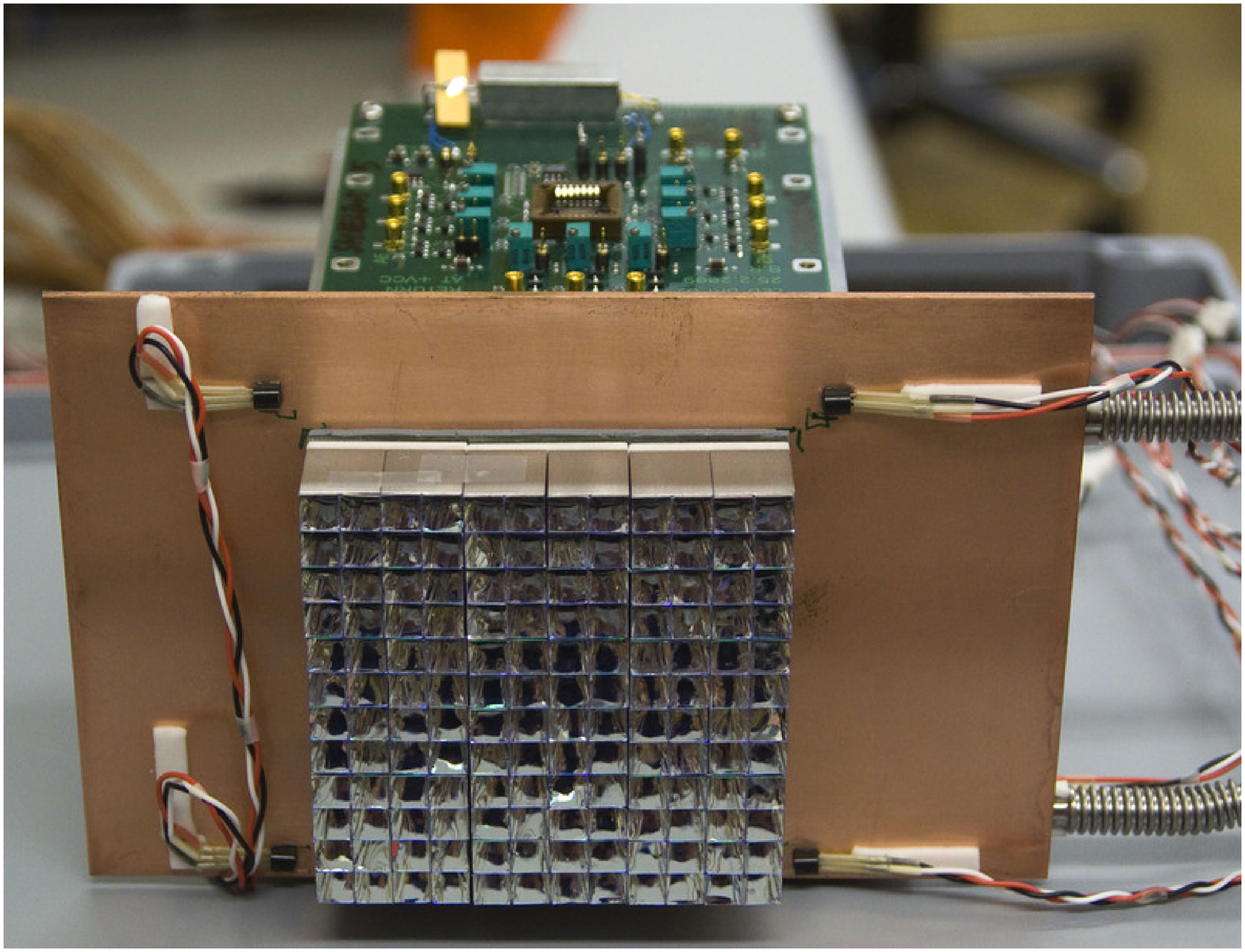}
    \hfill
    \includegraphics[height=3.7cm]{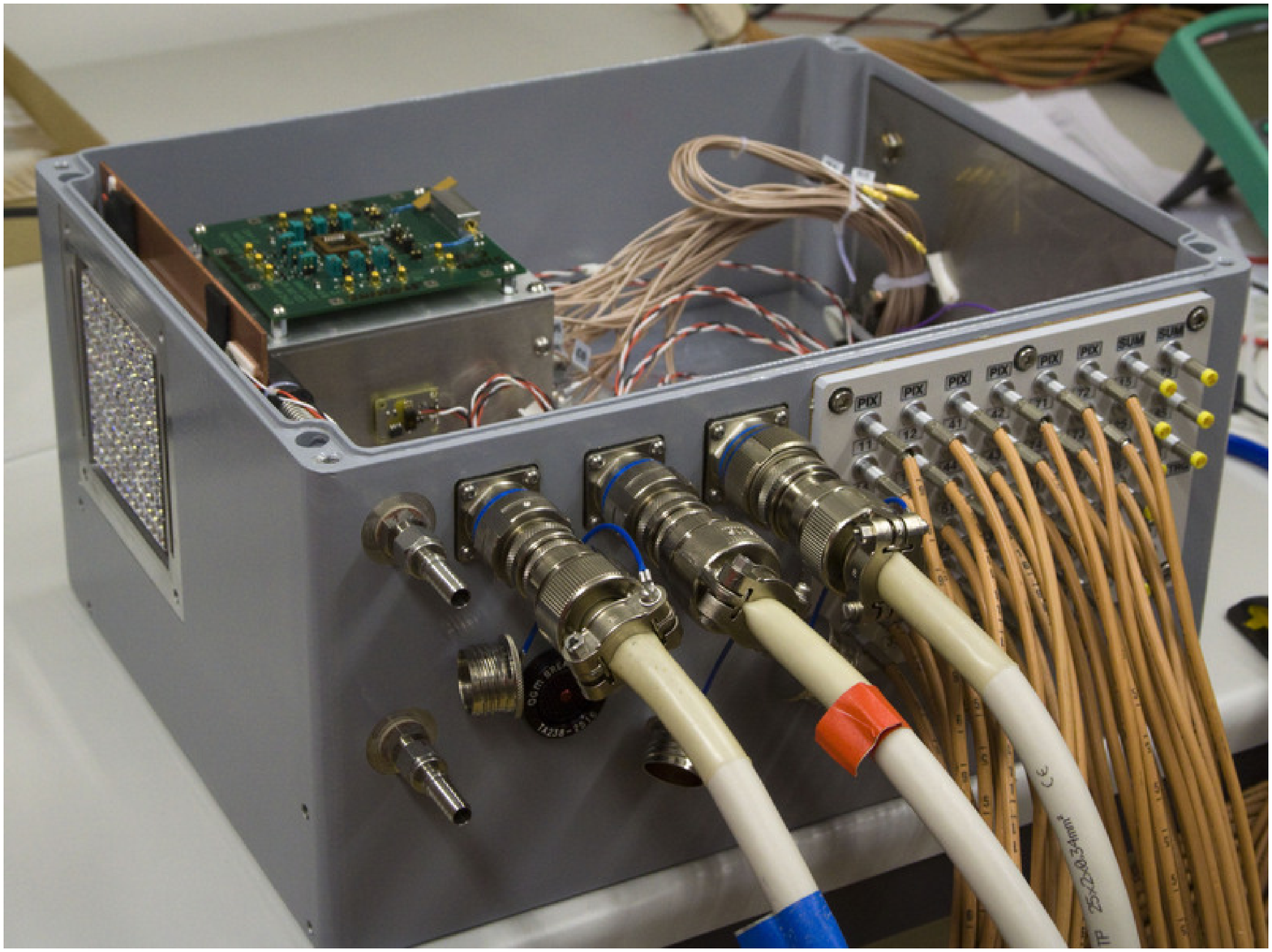}
    \caption{Photographs taken during the camera assembly. Left: carrier PCB
      with 16 G-APDs ($\mathrel{\widehat{=}}$~4~pixel) attached to the front,
      and the three front-end electronics boards attached to the back; a block
      of four light collectors ($\mathrel{\widehat{=}}$ 1 pixel) is mounted 
      separately for demonstration
      purposes. Middle: fully assembled module including cooling plate. Right:
      integration into water-tight box; signal and voltage supply cables
      attached.}
    \label{fig:camera}
  \end{center}
\end{figure}

The signals from four G-APDs are summed up in front-end electronics boards
(FEB) \cite{roeser08}, resulting in 36 quadratic pixels of 14.4\,mm $\times$
14.4\,mm size. These boards also perform a signal shaping and amplification on
the analog level (4\,mV voltage output per \textmu A current input, pulse
decay times $<$ 10\,ns). Each FEB comprises twelve channels, with a power
consumption of 150\,mW per channel. Figure~\ref{fig:camera} (left) shows a
photograph of the camera module during assembly, including the FEBs. Also
16 of the 144 G-APDs are visible, and a block of four light collectors which
are usually mounted on top of the sensors. Such a block corresponds to one
pixel (readout channel).

In order to keep the gain of the G-APDs stable in case of changes of the
ambient temperature and NSB conditions\footnote{The NSB light produces a
  permanent current in a serial resistor of the readout electronics and
  thereby reduces the effective bias voltage, which is supplied through the
  FEBs.}, a bias voltage feedback system has been foreseen. A calibration
signal from a pulsed and temperature-stabilized light emitting diode can be
monitored continuously and, if a change in amplitude is detected on a certain
channel, the voltage of the corresponding pixel is modified to readjust the
amplitude. Special power supply modules have been developed for this task
\cite{commichau08}, which can communicate with the camera control software
(see also section~\ref{sec:daq}) via an USB interface and allow the bias
voltage of each pixel to be regulated individually\footnote{Since four G-APDs
  are combined in one pixel, they have been selected such that their nominal
  operating voltages are within $\pm 10$\,mV. This translates into gain
  variations of a few percent \cite{hamamatsu08}.}. In addition, a water
cooling system has been installed. An external cooling unit provides water of
an adjustable temperature, which is pumped through copper tubes soldered onto
a copper plate. The G-APDs are coupled to this plate by a thermally conductive
but electrically insulating filling material. Several temperature sensors and
one humidity sensor are used to monitor the conditions inside the box and on
the cooling plate, as can be seen from figure~\ref{fig:camera} (middle). The
cooling system is not obligatory to operate the sensors, but can be used
optionally. The whole camera module is mounted inside a water-tight box (see
figure~\ref{fig:camera}, right), which has a protection window, a separate
shutter (not shown in the photograph) and connectors for the signal and
voltage supply cables.

\section{Trigger and data acquisition system}
\label{sec:daq}

The analog signals are transferred from the camera box to the counting room by
means of 20\,m long coaxial cables. They are fed into linear fan-out modules
(NIM standard)~\cite{roeser09}. One module comprises twelve sub-units, each
with one input channel and two normal and one inverted output channels. One of
the (positive) outputs is connected to the data acquisition (DAQ) system,
while the inverted (negative) signal is used for triggering purposes (see
figure~\ref{fig:schematic}). The trigger logic consists of a CAEN V812 VME
board where a majority coincidence of the innermost $16$ pixels is
formed. Individual trigger thresholds can be set via the VME bus.

The DAQ itself is based on the DRS2 (Domino Ring Sampling) chip \cite{ritt04}
containing ten analog pipelines of $1024$ capacitive cells. Signal sampling is
performed with a frequency of $0.5$--$4.5$\,GHz, generated on-chip by a series
of inverters. Such high sampling frequencies are desired for an IACT, since
excellent timing significantly improves the reconstruction of the properties
of the primary particle \cite{aliu09}. Each pipeline is read out at $40$\,MHz
and externally digitized by a multiplexed $12$\,bit flash ADC
(analog-to-digital converter). The DRS2 chips are mounted on mezzanine cards,
which are hosted by VME boards (two chips per card and two cards per VME
board). In this design, eight channels of each chip are available for external
signals (see also \cite{ritt04}). A single-board computer is used as VME
controller, with an attached hard disk for data storage. The DAQ and bias
voltage control software are also running on this computer. For a trigger rate
of $10$\,Hz, typical data rates are of the order of $1$\,MByte/s.

Sampling the signal with the ring sampler at frequencies in the GHz range
allows a precise photon arrival time measurement, provided the so-called
fixed-pattern aperture jitter of the DRS2 chip is corrected. This relatively
large, but systematic and gradual spread in its bin widths results from the
manufacturing process. As the occurrence of a trigger is random with respect
to the physical pipeline, the time measured between the trigger and the
signal, being the sum of the widths of the involved bins, is also randomized
to some degree. At 2\,GHz sampling, this jitter can amount to about 5\,ns and
can have a complicated distribution, depending on the distance between trigger
and signal. It can be corrected by calibrating with a high-frequency signal
from a precision generator. The procedure involves measuring the period of
this signal with the DRS2, and stretching or shrinking of the bin widths
within that particular period to match the generator output. As the revolution
frequency of the domino wave is phase locked to a stable oscillator, the
revolution time (the sum over all 1024 individual bin widths) itself is
fixed. Therefore, the inverse correction is applied to all bins outside the
current period. Doing this for all periods of a sampled waveform and
iteratively for many waveforms with random phase relative to the domino wave
converges towards the required correction values. The effectiveness of the
jitter correction is shown in figure~\ref{fig:jitter} (left). The histograms
show the time distributions of the rising edge of a pulse relative to the
trigger signal with and without the correction at 2\,GHz sampling
frequency. Both the (square) test-pulse and the trigger were derived from a
single output of a pulse generator. The correction values were determined in
this case using a 200\,MHz rectangular signal. After applying them, the pulse
time distribution has a width of 390\,ps root-mean-square.

\section{Measurement setup and data taking parameters}
\label{sec:setup}

\begin{figure}[t]
  \begin{center}
    \includegraphics[height=6.2cm]{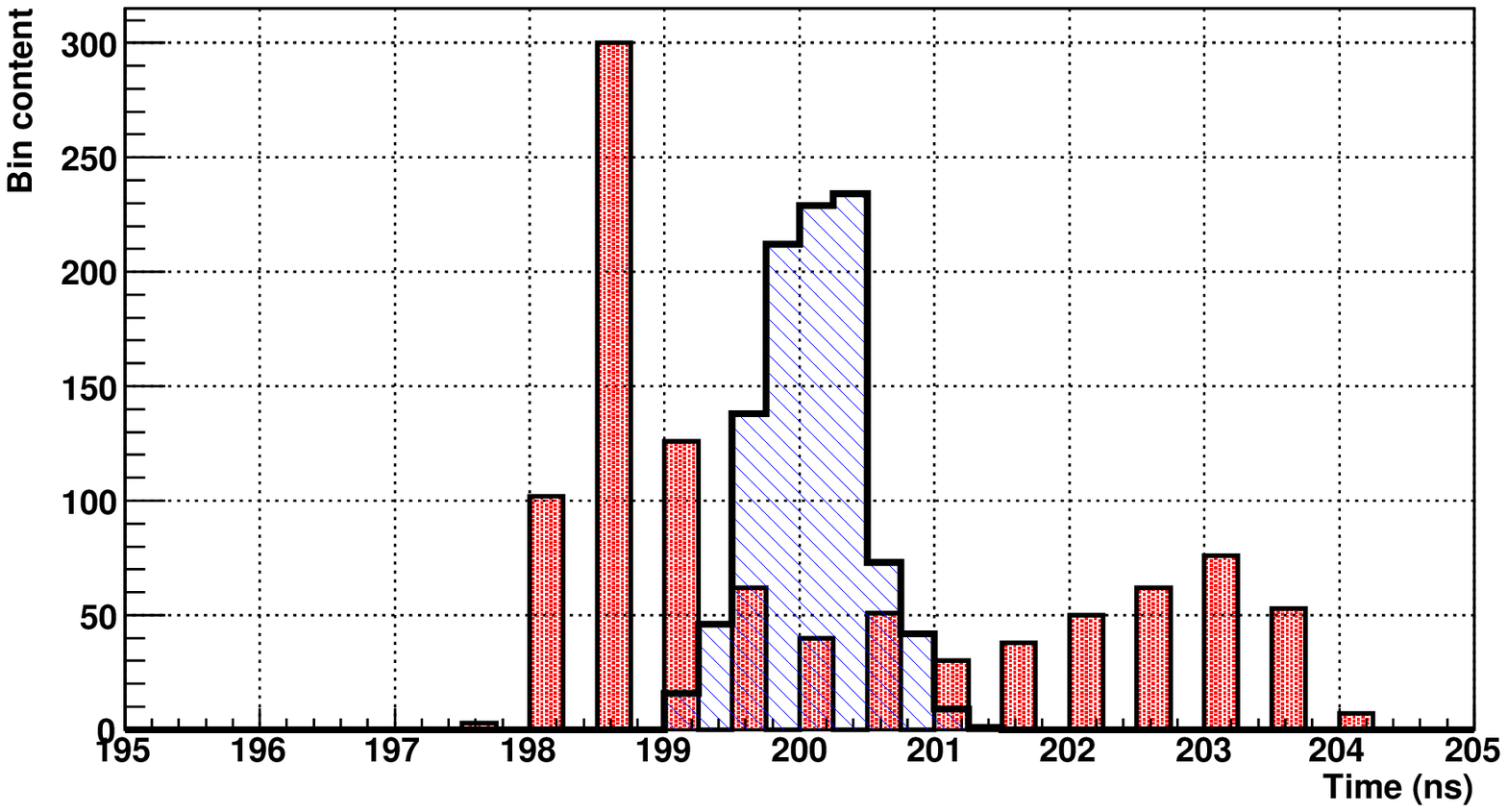}
    \hfill
    \includegraphics[height=6.2cm]{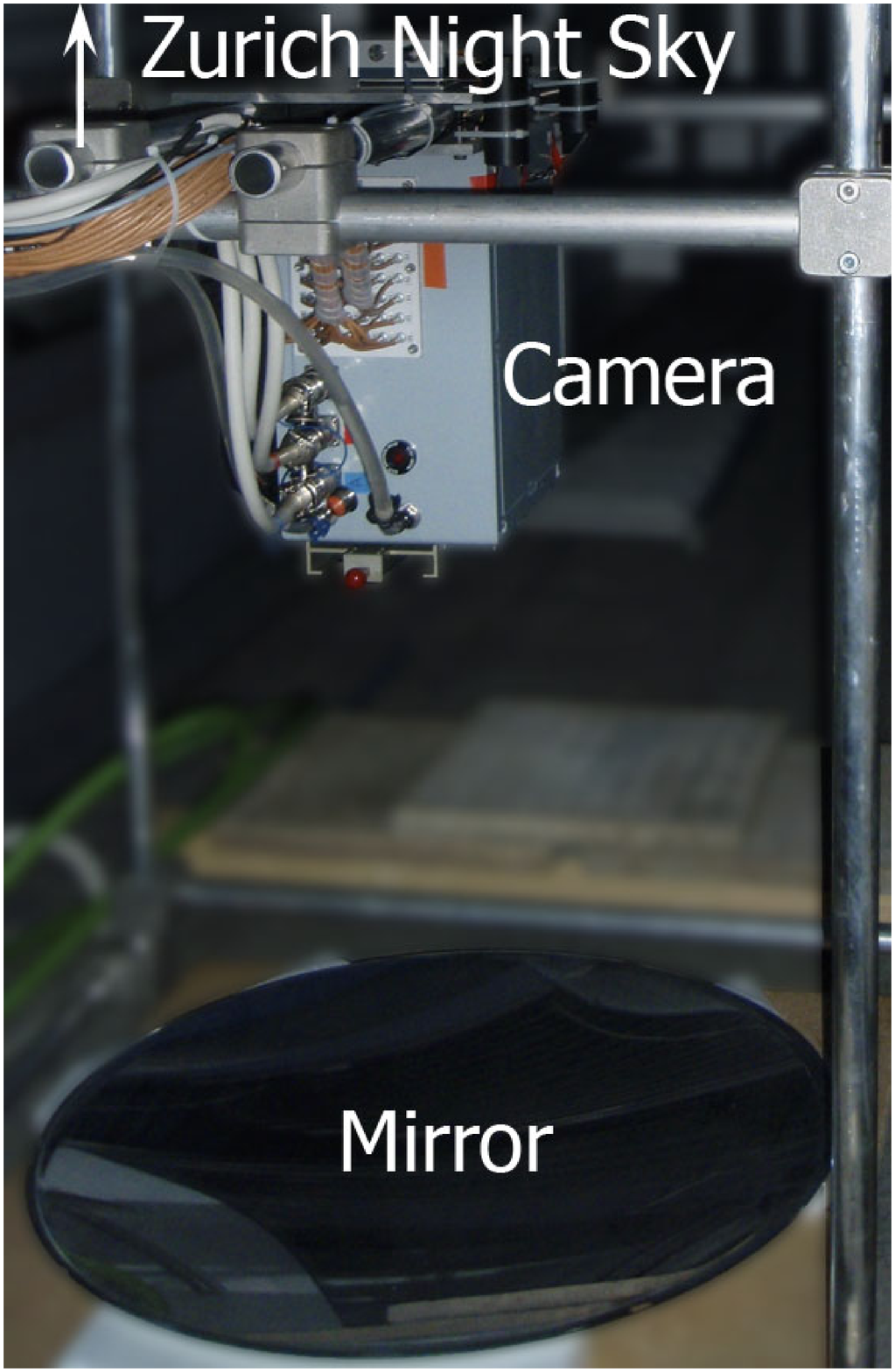}
    \caption{Left: time distribution of square pulses sampled
      at 2\,GHz with the DRS2
      chip without (red filled histogram) and with jitter correction (blue
      shaded). The root-mean-square of the latter is 390\,ns. Right:
      photograph of the experimental setup to record cosmic-ray induced air
      showers (see also text).}
    \label{fig:jitter}
  \end{center}
\end{figure}

In figure~\ref{fig:jitter} (right) the experimental setup used for the
detection of cosmic air showers is shown. The camera was mounted in the focus
of a zenith-pointing spherical mirror (90\,cm diameter). The focal length of
the mirror was 80\,cm which, taking into account the pixel dimensions,
translates into a field of view of $1.0^{\circ}$ per pixel. The measurements
were performed during the night of July 2, 2009, near the city center of
Zurich. Thus the NSB from surrounding buildings and also from partial
moonlight was rather high. It was determined with an external sky quality
meter (Unihedron) to be 300\,MHz per G-APD (1.2\,GHz per pixel). During the
measurements, which took about an hour, the outdoor temperature was around
$22^{\circ}$C. The G-APD plane was cooled to $18^{\circ}$C and the bias
voltage for all pixels set to the nominal values for this temperature ($\sim
400$\,mV lower than for operation at $25^{\circ}$C). Under these stable
conditions, no bias voltage feedback had to be used.

A sampling frequency of 2\,GHz was set for the DRS2 chips. The trigger
threshold was adjusted for each of the innermost 16 pixels, such that single
pixel trigger rates of 1--3\,kHz were obtained for an opened camera
shutter. This was achieved by applying discriminator thresholds of $40 \pm
4$\,mV to the analog signals\footnote{The thresholds had to be exceeded for at
  least 4\,ns to fulfill the trigger requirement. A laboratory calibration,
  though without NSB light, showed that the analog readout features 7\,mV
  signal amplitude per fired G-APD cell (approximately equivalent to the
  number of incoming photons per pixel multiplied with the PDE of the
  G-APDs).}. A majority coincidence of four channels above threshold within a
time window of 20\,ns was furthermore required. Under these circumstances, the
DAQ system recorded data with an average trigger rate of 0.02\,Hz (also
including noise triggers). The latency of the whole system was about 350\,ns.

\section{Results and discussion}

\begin{figure}[t]
  \begin{center}
    \includegraphics[width=0.78\textwidth]{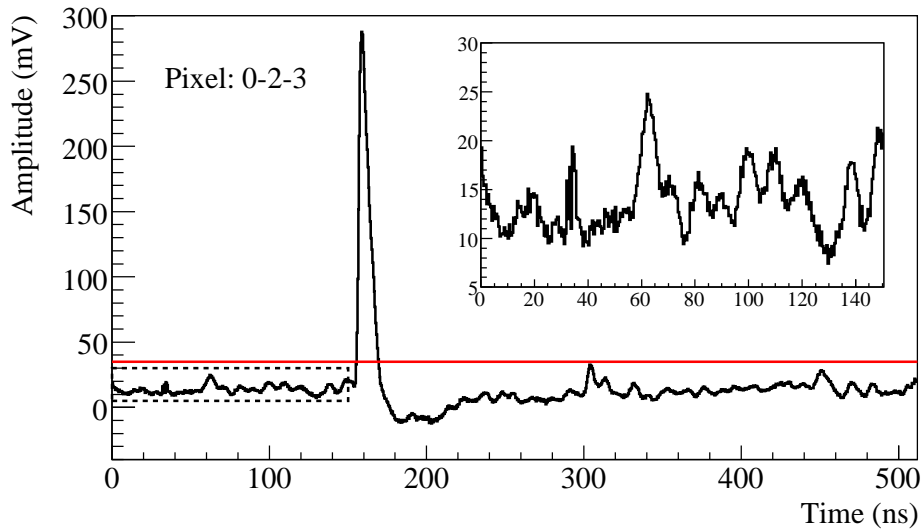}
    \caption{Raw data of one pixel recorded with a sampling frequency of
      2\,GHz; full DRS2 pipeline shown. A Cherenkov-light signal can be seen
      at about 160\,ns. The red solid line indicates the trigger
      threshold. The inset shows a zoom to the data from 0--150\,ns as marked
      by the dashed-line~box.}
    \label{fig:signal}
  \end{center}
\end{figure}

For the offline analysis, events corresponding to Cherenkov light flashes have
been filtered out from the recorded data. Several neighboring channels
containing a clear signal, at the time position expected from the trigger
latency, have been required. This is demonstrated in figure~\ref{fig:signal},
which presents the raw data for a certain pixel (event \#16 of run \#207). The
large plot shows the full DRS2 pipeline with a Cherenkov-light signal at about
160\,ns, while the inset presents a zoom to the data between 0 and 150\,ns. In
the latter, the fluctuating signals from NSB photons are visible, which are
very frequent and therefore pile up. The red line indicates the trigger
threshold\footnote{Translated to the digitized sample amplitudes. Please note
  that the signal is attenuated in front of the DRS2 chip to match its dynamic
  range.}. From the data analysis it has been estimated that the trigger rate
for air shower events was about 0.01\,Hz.

\begin{figure}[t]
  \begin{center}
    \includegraphics[width=0.49\textwidth, clip]{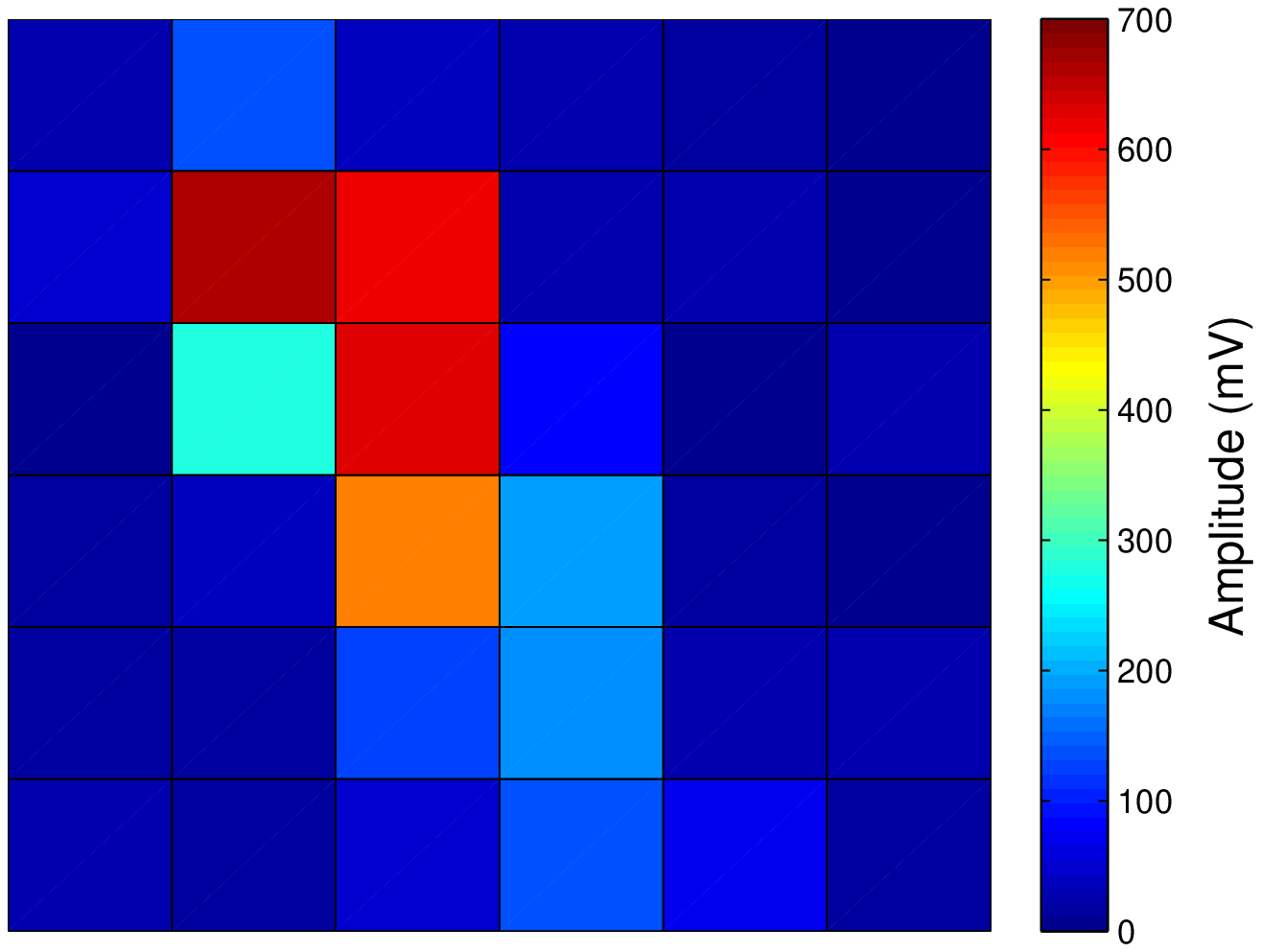}
    \hfill
    \includegraphics[width=0.49\textwidth, clip]{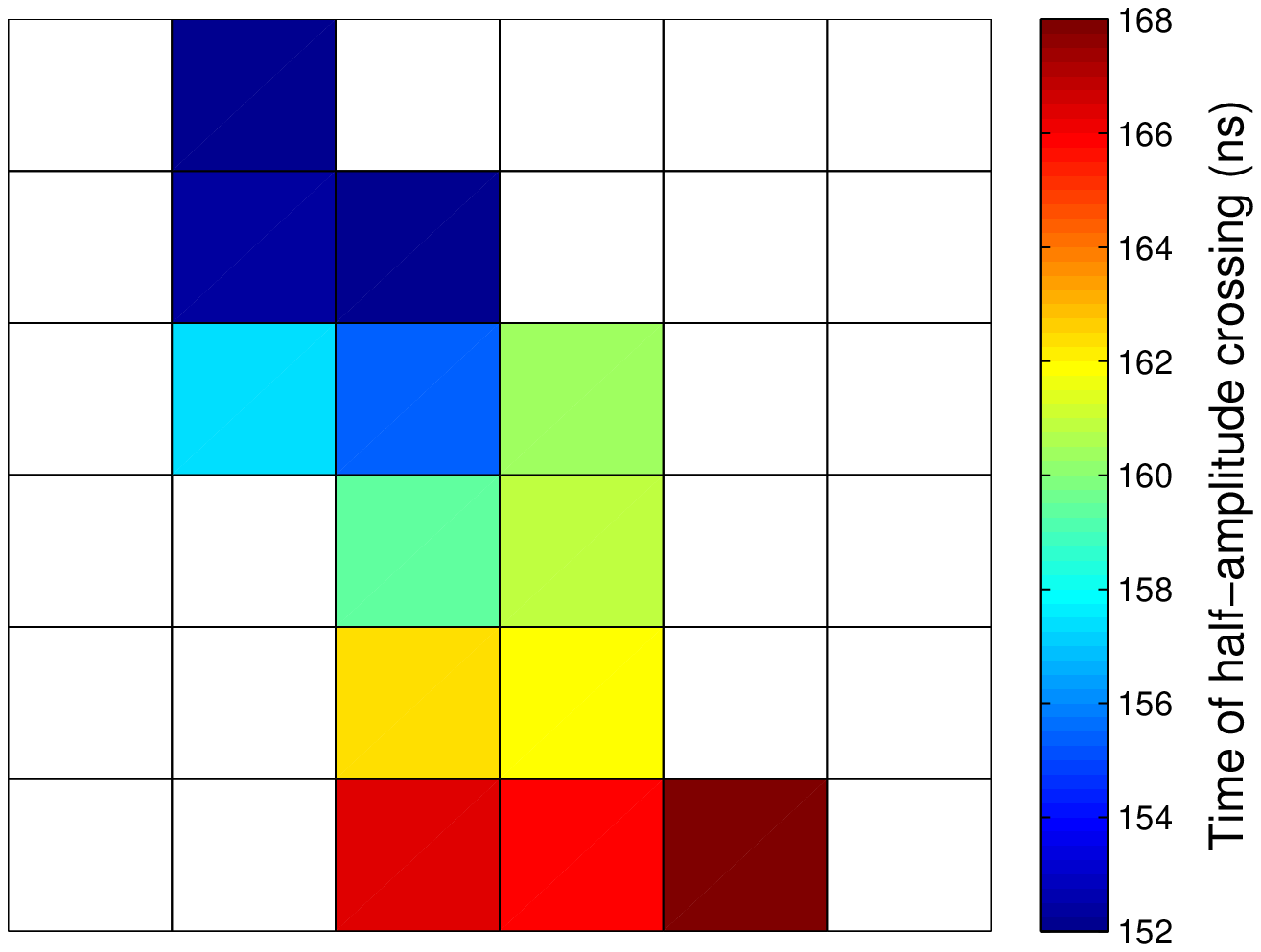}
    \caption{Air shower image recorded with the G-APD camera (event \#16 of
      run \#207); the signal shown in figure~\protect\ref{fig:signal} belongs
      to the pixel in the second column from the left, third row from the
      top. Left: intensity distribution over the 6 pixels $\times$ 6
      pixels. Right: corresponding signal arrival time distribution (see
      text).}
    \label{fig:shower}
  \end{center}
\end{figure}

\begin{figure}[t]
  \begin{center}
    \includegraphics[width=0.49\textwidth, clip]{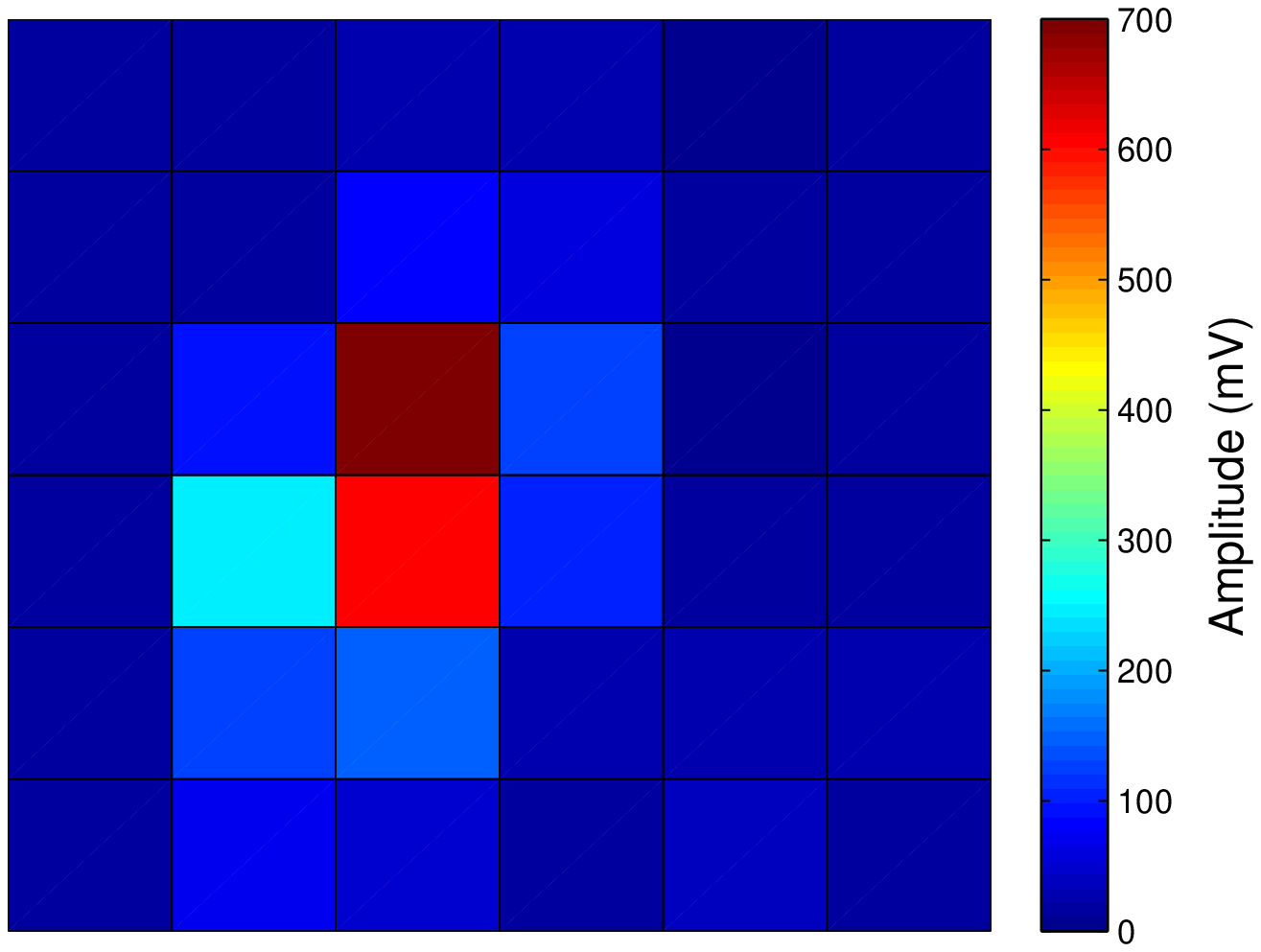}
    \hfill
    \includegraphics[width=0.49\textwidth, clip]{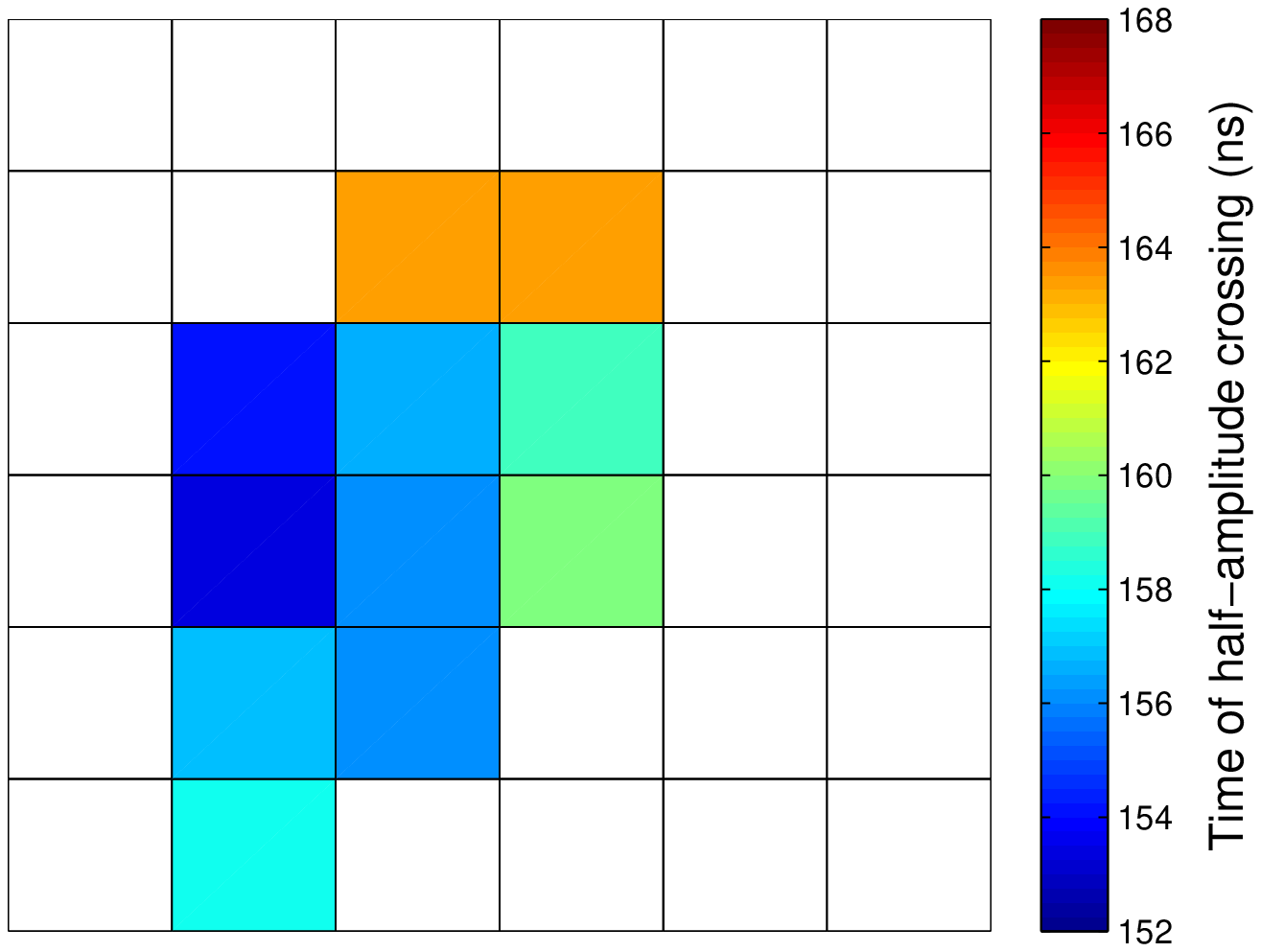}
    \caption{Further air shower image recorded with the G-APD camera (event
      \#14 of run \#206). Left: intensity distribution over the 6 pixels
      $\times$ 6 pixels. Right: corresponding signal arrival time distribution
      (see text). The color scales are the same as in
      figure~\protect\ref{fig:shower}.}
    \label{fig:shower2}
  \end{center}
\end{figure}

In figure~\ref{fig:shower} (left) the intensity distribution for the whole
event introduced above is plotted. The maximum sample amplitude is shown on
each of the 36 pixels, searched for within a time window of 100\,ns around the
expected signal position. Each amplitude has been corrected for the DRS2
baseline, evaluated event-by-event from the data samples recorded before the
signal arrival (see figure~\ref{fig:signal}). On the right side of
figure~\ref{fig:shower}, the jitter corrected (cf. section~\ref{sec:daq})
arrival time of the signals is presented\footnote{Defined as the time of the
  last sample where the signal is below $50\%$ of its amplitude (no
  interpolation).}. Only pixels with a signal amplitude above 60\,mV have been
used for the timing calculation, the others are displayed in white color. A
clear shower development is apparent for this event, starting in the top left
corner of the display and extending to the bottom right corner within almost
20\,ns. Figure~\ref{fig:shower2} shows an event with a less extended time
distribution (event \#14 of run \#206). Especially the core pixels with the
largest amplitudes are within a few ns. The two outer pixels with a time
offset of 3--4\,ns compared to the core pixels are likely due to a sub-shower.

Taking into account that the camera covers a field of view of $6^{\circ}$ in
both dimensions (see also section~\protect\ref{sec:setup}), such events
certainly come from air showers induced by very high energetic cosmic-ray
particles. Dedicated simulations have been carried out\footnote{The air shower
  simulation using CORSIKA \cite{heck98}, the detector simulation and data
  analysis using the MARS package (CheObs edition) \cite{bretz09}.},
concluding that the experimental setup described in this paper has an energy
threshold of several TeV. Primary protons with off-axis angles up to
5$^{\circ}$ have been simulated\footnote{Protons compose the largest fraction
  of the cosmic rays. Since in this case the mirror axis points up vertically
  to the sky, the off-axis angle coincides with the zenith angle.}, and photon
arrival time distributions consistent with the event presented in
figure~\ref{fig:shower2} have been observed. The event shown in
figure~\ref{fig:shower} most probably corresponds to a primary particle with
an off-axis angle of 10$^{\circ}$ or larger. Because of the comparatively long
trigger coincidence window, it was possible to trigger such events.

\section{Conclusion}

A 36-pixel prototype G-APD camera for Cherenkov astronomy has successfully
been constructed and commissioned. A DAQ system based on the DRS2 chip has
been set up. In a self-triggered mode, images of air showers induced by cosmic
rays have been recorded. For the first time, this has been achieved with a
complete G-APD camera. Stable operation at room temperature and under high NSB
light conditions is possible. In summary, these photo-sensors have proven to
fulfill the requirements of IACT applications, with the potential to replace
or complement PMTs for future projects like the planned Cherenkov Telescope
Array (CTA) \cite{cta}.

\begin{acknowledgments}
  This work is supported by ETH research grants 0-43391-08 and 0-20486-08.
\end{acknowledgments}

\end{document}